\documentclass[12pt]{article}
\usepackage{amsmath,amssymb, graphicx}

\numberwithin{equation}{section}

\def\mydate{September 18,  2007}
\def\ignore#1{{}}

\newcounter{sxn}

\newcounter{axn}

\date{}

\newdimen\mybaselineskip
\renewcommand{\baselinestretch}{1.25}

\renewcommand{\thefootnote}{\arabic{footnote}}

\newcommand{\beeq}{\begin{equation}}
\newcommand{\eneq}{\end{equation}}
\newcommand{\beqn}{\begin{eqnarray}}
\newcommand{\eeqn}{\end{eqnarray}}

\def\mybig{\displaystyle \strut }

\def\dd{\partial}
\def\la{\raise.16ex\hbox{$\langle$}\lower.16ex\hbox{}  }
\def\ra{\, \raise.16ex\hbox{$\rangle$}\lower.16ex\hbox{} }
\def\go{\rightarrow}

\def\onehalf{ \hbox{${1\over 2}$} }

\def\Tr{{\rm Tr \,}}

\def\eff{{\rm eff}}

\def\cL{{\cal L}}

\def\diag{{\rm diag ~}}

\def\hp{{\hat p}}
\def\hq{{\hat q}}

\def\ep{\epsilon}
\def\psibar{ \psi \kern-.65em\raise.6em\hbox{$-$} }
\def\psibarl{ \psi \kern-.65em\raise.6em\hbox{$-$} \lower.6em\hbox{} }

\def\sp{\hat p \kern-.5em{ /} \kern+.2em}
\def\sq{\hat q \kern-.5em{ /} \kern+.2em}
\def\spp{p \kern-.5em{ /} \kern+.2em}
\def\sqq{q \kern-.5em{ /} \kern+.2em}

\def\myfrac#1#2{{\mybig #1\over \mybig #2}}

\begin{document}
\thispagestyle{empty}

\baselineskip=12pt

{\small \noindent \mydate    \hfill OU-HET 585/2007}

\rightline{\small  KOBE-TH-07-07}

\rightline{\small  TU-797}
\rightline{\small  SISSA  63/2007/EP}


\baselineskip=35pt plus 1pt minus 1pt

\vskip 2.5cm

\begin{center}
{\Large \bf  Two loop finiteness of  Higgs mass and potential}\\
{\Large \bf in the gauge-Higgs unification}\\

\vspace{2.5cm}
\baselineskip=16pt  

{\def\thefootnote{\fnsymbol{footnote}}
\bf 
Y.\ Hosotani$^1$\footnote[1]{hosotani@phys.sci.osaka-u.ac.jp},
N.\ Maru$^2$\footnote[2]{maru@people.kobe-u.ac.jp},
K.\ Takenaga$^3$\footnote[3]{takenaga@tuhep.phys.tohoku.ac.jp}
\ignore{
Yutaka Hosotani$^1$\footnote[1]{hosotani@phys.sci.osaka-u.ac.jp},
Nobuhito Maru$^2$\footnote[2]{maru@people.kobe-u.ac.jp},
Kazunori Takenaga$^3$\footnote[3]{takenaga@tuhep.phys.tohoku.ac.jp} 
and\\
}
and
Toshifumi  Yamashita$^4$\footnote[4]{yamasita@sissa.it}
}\\
\vspace{.3cm}
$^1${\small \it Department of Physics, Osaka University,
Toyonaka, Osaka 560-0043, Japan}\\
$^2${\small \it Department of Physics, Kobe University,
Kobe 657-8501, Japan}\\
$^3${\small \it Department of Physics, Tohoku University,
Sendai 980-8578, Japan}\\
$^4${\small \it Scuola Internazionale Superiore di Studi Avanzati,
                via Beirut 2-4, I-34014 Trieste, Italy}\\
\end{center}

\vskip 2.0cm
\baselineskip=20pt plus 1pt minus 1pt

\begin{abstract} 
The zero mode of an extra-dimensional component of gauge potentials 
serves as a 4D Higgs field in the gauge-Higgs unification.
We examine QED on $M^4 \times S^1$ and determine  the mass and potential 
of a 4D Higgs field (the $A_5$ component) at the two loop level 
with  gauge invariant reguralization. 
It is seen that the mass  is free from divergences and  independent 
of the renormalization scheme. 
\end{abstract}


\newpage
\section{Introduction}

In the standard model of electroweak interactions the Higgs boson is vital 
to induce the electroweak symmetry breaking.  It is one of the major goals
in particle physics to discover the Higgs boson  in the coming years. 
Its  mass squared $m_H^2$, in general, 
acquires $O(\Lambda^2)$ radiative corrections 
where $\Lambda$ is a cutoff scale which is as large as 
 $10^{16} \,$GeV in grand unified theories.
In order to have $m_H = O(100) \,$GeV, unnatural fine-tuning of 
parameters of the theory is demanded.   

The supersymmetry naturally  solves this gauge hierarchy problem to
push down $\Lambda$ to the TeV scale.  It serves as a leading candidate
for a model beyond the standard model, and is under intensive study.
There are alternative scenarios to have a naturally light Higgs boson,
among which is the gauge-Higgs unification.\cite{Fairlie1}-\cite{Oikonomou}
A 4D Higgs field is identified
with a part of the extra-dimensional component of gauge potentials.
When the extra-dimensional space is not simply connected, there appears
a Wilson line phase $\theta_H$, an analogue  of the 
Aharonov-Bohm phase in quantum mechanics.  The 4D Higgs field
is nothing but a field describing four-dimensional fluctuations of $\theta_H$.
At the tree level the Higgs field appears massless, reflecting the nature of the 
Aharonov-Bohm phase.  At the quantum level the effective potential
for the Wilson line phase, $V_\eff (\theta_H)$, is generated radiatively.
It has been shown long ago that $V_\eff (\theta_H)$ on $M^4 \times S^1$ 
and  the Higgs mass $m_H$ are finite 
at the one loop level.\cite{YH1, YH2, McLachlan1}

This has significant relevance in the context of the gauge-Higgs unification.
Although higher dimensional gauge theory is not renormalizable, 
the 4D Higgs mass can be predicted to be a finite value without afficting from 
the problem of  of divergences.  As the Higgs field is associated with 
the nonlocal Wilson line phase,  it is commonly said 
that the Higgs mass
remains finite to all order as  no gauge-invariant local counter term
can be written.  It is not quite clear, however,  whether this argument applies
to non-renormalizable theories like the one under consideration.  It is desirable
to have explicit  evaluation and confirm the finiteness of $m_H$ 
beyond one loop.

There have been significant advances in 
the gauge-Higgs unification in the electroweak theory  
 in the last couple of years.  In the early stage
unification on orbifolds $M^4 \times (S^1/Z_2)$ and $M^4 \times (T^2/Z_2)$ 
was pursued with chiral fermions.\cite{Pomarol1}-\cite{HMOY} 
It has been recognized  
that unification on warped spacetime such as the Randall-Sundrum 
spacetime works much better for having phenomenologically viable 
models.\cite{Pomarol2}-\cite{Wagner1}
In all of such aspects  as the Higgs mass, the Kaluza-Klein mass scale, 
the gauge self-couplings, and the Weinberg angle,  the unification 
in the Randall-Sundrum spacetime gives natural consistent results.
Suppression of the Higg-gauge couplings and Yukawa 
couplings  has been predicted, which can be tested at LHC.\cite{SH1, HS2}
Furthermore it has been shown recently that the gauge-Higgs 
unification in the Randall-Sundrum spacetime is dual to the theory of
holographic pseudo-Goldstone boson.\cite{Falkowski}-\cite{Sakamura}

In view of these developments it is appropriate and necessary
to strengthen and confirm the statement that the Higgs mass 
remains finite beyond one loop.\cite{Gersdorff}-\cite{YHfinite}
Its finiteness  has been investigated  in the lattice 
simulation on orbifolds as well.\cite{Irges}
The calculability  of the $S$ and $T$ parameters 
in the electroweak theory has been  discussed at the one loop level.\cite{LM1}

Evaluation of the Higgs mass at the two loop 
level is formidable in non-Abelian gauge theory.  To get insight in the
problem it is instructive to examine, as the first step, 
QED on $M^4 \times S^1$ in which the zero 
mode of the extra-dimensional component $A_5$ mimics 
the 4D Higgs boson.  It is  called as a Higgs boson in the
present paper.  
 
To evaluate the Higgs mass $m_H$ at the two loop level,
renormalization at the one loop level must be  taken into 
account in due course.  
Two loop evaluation of the Higgs mass in QED on $M^4 \times S^1$ has been
previously attempted by Maru and Yamashita\cite{MY1}, where
the vacuum polarization tensors $\Pi^{MN}$ are evaluated near $\theta_H =0$ 
for the zero-modes, without paying serious attention to  the reguralization.
In this article 
we evaluate both the effective potential $V_\eff (\theta_H)$ and
the vacuum polarization tensors  in 
renormalized perturbation theory in the dimensional regularization,
maitaining the gauge invariance and the Ward-Takahashi identities.
The computation is carried out with an arbitrary value of $\theta_H$ 
as a background.  The effective potential $V_\eff (\theta_H)$  is found to
be minimized at $\theta_H = \pi$, and therefore the vacuum polarization
tensors $\Pi^{MN}$ at $\theta_H = \pi$ become relevant for determining $m_H$.

The paper is organized as follows.  In the next section  renormalized 
perturbation theory for QED in $M^4 \times S^1$ is developed and
renormalization conditions are given.
In Section 3 the effective potential $V_\eff (\theta_H)$ is evaluated at the
 two loop level.  Relevant integral-sums are evaluated in Appedix A.
In Section 4 the vacuum polarization tensors $\Pi^{MN}$ are determined
at the one loop level.  Details of the computaion are given in Appendix B.
With these results the Higgs mass is determined at the two loop level in Section 5.
It is seen that the Higgs mass thus evaluated is independent of 
the renormalization scheme.
Section 6 is devoted to a brief summary and discussions.

\section{QED in $M^4 \times S^1$}

The model we analyze is QED defined in five-dimensional spacetime 
where the fifth dimension is a circle $S^1$ with a radius $R$.
For the sake of simplicity  we introduce only one fermion $\psi$
with a mass $m$.  
Both the gauge potential $A^M$ (the photon field) and $\psi$ are
taken to be periodic.  Renormalization, at least at the two loop level, 
is done with the standard renormalization procedure.  Renormalized 
fields are defined by $A^{(0)}_M = Z_3^{1/2} A_M$ and 
$\psi^{(0)} = Z_2^{1/2} \psi$.  The renormalized coupling constant
is defined by $e^{(0)} = Z_1 Z_2^{-1} Z_3^{-1/2} e$.  The renormalized mass
of $\psi$ is given by $m^{(0)} = m + \delta m$.  Here quantities with
superscript $(0)$ denote bare quantities.   Renormalization conditions 
for $Z_1, Z_2, Z_3$ and $\delta m$ are specified below.

We develop renormalized perturbation theory around  the non-vanishing 
Wilson line phase $\theta_H$.  The Lagrangian density is 
given by
\beqn
&&\hskip -1cm
\cL = - \frac{1}{4} F_{MN} F^{MN} - \frac{1}{2} (\dd_M A^M)^2 
+ \psibar (i \gamma^M D^c_M -m ) \psi  - e A_M \psibar \gamma^M \psi \cr
\noalign{\kern 10pt}
&&\hskip -.2cm
- \frac{1}{4} \delta_3 F_{MN} F^{MN} 
+ \psibar (\delta_2 i \gamma^M \dd_M - \delta_m) \psi 
- e \delta_1   A_M \psibar \gamma^M \psi \ ~~.
\label{Lagrangian1}
\eeqn
Here $D^c_M = \dd_M -  \delta_{M,5} ie A_5^c$ where
$e  A_5^c = \theta_H / 2\pi R$ and  the metric
is $\eta_{MN} = \diag (1, -1,-1,-1,-1)$  
Counter terms are defined as
$\delta_k = Z_k -1$ and $\delta_m = Z_2 m^{(0)} - m$.  
We renormalize such that the Ward-Takahashi identity $Z_1=Z_2$ 
is preserved  so that
\beeq
\theta_H = e \int_0^{2\pi R} dy \, A_5 
= e^{(0)} \int_0^{2\pi R}  dy \,  A_5^{(0)} ~.
\label{Wphase1}
\eneq
In other words the Wilson line phase is not renormalized.

On $M^4 \times S^1$ the fifth component of a momentum $p^5$ is 
discretized.   With $\theta_H \equiv 2\pi a \not= 0$, 
$p^5 = (n-a)/R$ for  $\psi$ and $p^5 = n/R$ for $A_M$ where
$n$ is an integer.  We denote a five-momentum by 
$\hp^M = (p^\mu, p^5)$ $(\mu=0 \sim 3)$.   
Let us denote  the sum of all 1-particle-irreducible diagrams
for the fermion propagator by $\Sigma (\hp \, ; a , R)$, that for the photon 
propagator by $\Pi^{MN} (\hp \, ; a , R)$, and the amputated fermion 
vertex function by $\Gamma^M (\hp, \hp{\,}' \, ; a , R)$. 
In the $R \go \infty$ limit these functions take 5D Lorentz covariant
form, and are denoted with a superscript $(0)$;
\beqn
\Sigma (\hp \, ; a , R) &=& \Sigma^{(0)} (\sp ) 
    + \Sigma^{(1)} (\hp \, ; a , R) ~~, \cr
\noalign{\kern 5pt}
\Pi^{MN} (\hp \, ; a , R) &=& \Pi^{(0) MN} (\hp ) 
    + \Pi^{(1) MN} (\hp \, ; a , R) ~~, \cr
\noalign{\kern 5pt}
\Gamma^{M} (\hp, \hp{\,}' \, ; a , R) &=& \Gamma^{(0) M} (\hp, \hp{\,}' ) 
    + \Gamma^{(1) M} (\hp, \hp{\,}' \, ; a , R) ~~, 
\label{invariant1}
\eeqn
where $\sp = \hp^M \gamma_M$ and 
$\Sigma^{(0)} (\sp ) = \lim_{R\go \infty} \Sigma (\hp \, ; a , R)$ etc..
The vacuum polarization tensors $\Pi^{MN} (\hp \, ; a , R)$ can be 
expressed in terms of two invariant functions $\Pi (\hp \, ; a , R)$ and
$F (\hp \, ; a , R)$.   The current conservation
$\hp_M \Pi^{MN}  (\hp \, ; a , R) = 0$ implies that
\beqn
&&\hskip -1cm
\Pi^{\mu\nu} = ( \eta^{\mu\nu} \hat p^2 - p^\mu p^\nu) \, \Pi
 - \frac{p^\mu p^\nu}{p^2} \cdot (p^5)^2 \, F  ~~, \cr
\noalign{\kern 7pt}
&&\hskip -1cm
\Pi^{55} = - p^2 (\,  \Pi + F) ~~, \cr
\noalign{\kern 7pt}
&&\hskip -1cm
\Pi^{5\mu}= \Pi^{\mu 5} = - p^5 p^\mu  (\, \Pi + F) ~~,
\label{invariant2}
\eeqn
where $p^5 = n/R$ ($n$ : an integer). 
We remark that $F$ is finite and $\lim_{R\go \infty} F = 0$.
The divergent contributions appear only for 
$\Pi^{(0)} = \lim_{R \go \infty} \Pi$, which can be cancelled by the 
counter term $\delta_3$ in (\ref{Lagrangian1}).  
Also  $ \lim_{R \go \infty} \Pi^{MN} = 
( \eta^{MN} \hat p^2 - \hp^M \hp^N) \, \Pi^{(0)}(\hp^2)$.

The full photon propagators $D^{MN} (\hp)$ are given by
\beqn
&&\hskip -1cm
i D^{\mu\nu} = \frac{1}{ \hp^2 (1-\Pi)} 
\Big( \eta^{\mu\nu} - \frac{p^\mu p^\nu}{p^2} \Big)
+  \frac{p^\mu p^\nu}{(\hp^2)^2}  
\bigg\{ 1 - \frac{(p^5)^2}{p^2 (1- \Pi - F)} \bigg\} ~, \cr
\noalign{\kern 10pt}
&&\hskip -1cm
i D^{\mu 5} = - \frac{p^\mu p^5}{(\hp^2)^2}  
\frac{\Pi + F}{1- \Pi - F } ~, \cr
\noalign{\kern 10pt}
&&\hskip -1cm
i D^{5 5} = 
- \frac{p^2}{(\hp^2)^2 (1- \Pi - F )}  + \frac{(p^5)^2}{(\hp^2)^2}  ~.
\label{propagator1}
\eeqn
In particular, for the zero modes ($p^5=0$)
\beqn
&&\hskip -1cm
D^{\mu\nu} |_{p^5 = 0} = \frac{-i}{ p^2 (1-\Pi)} 
\Big( \eta^{\mu\nu} - \frac{p^\mu p^\nu}{p^2}  \Pi \Big)  ~, \cr
\noalign{\kern 10pt}
&&\hskip -1cm
D^{\mu 5} |_{p^5 = 0} =0~, \cr
\noalign{\kern 10pt}
&&\hskip -1cm
D^{5 5} |_{p^5 = 0} =  \frac{i}{p^2 (1- \Pi - F )}   ~,
\label{propagator2}
\eeqn
where $\Pi$ and $F$ are evaluated at $p^5 =0$.
It will be found that $\Pi$ and $F$ are expanded in $p^2$ as 
$\Pi  |_{p^5 = 0} = c_0 + \cdots$ and 
$F |_{p^5 = 0} = b_{-1} / p^2 + b_0 + \cdots$, respectively.  
Consequently the Higgs mass, or the mass of the zero mode of $A_5$,
defined in the $p^2$ expansion in the inverse propagator is given by
\beeq
m_H^2 = \frac{b_{-1}}{1 - c_0 - b_0} ~~.
\label{Higgs1}
\eneq
It slightly differs from the exact pole mass, but is convenient to relate to the
effective potential $V_\eff (\theta_H)$.  The difference between the 
two is small in the weak coupling $e_4^2/4\pi \ll 1$ where $e_4$ is
the four-dimensional gauge coupling $e_4^2 = e^2/ 2\pi R$.

There are two typical ways to impose renormalization conditions.  
The most convenient way is to impose them in the $R \go \infty$ limit,
namely in $M^5$. In terms of $\Sigma^{(0)}$,  $\Pi^{(0)}$ and
$\Gamma^{(0)}$ in (\ref{invariant1}) and   (\ref{invariant2}) 
and the five-dimensional fermion mass $m$ in $M^5$,  
renormalization constants $Z_j$ and $\delta m$ are fixed by
\beqn
&&\hskip -1cm
\Sigma^{(0)} (\sp=m)  = 0 ~~,  \cr
\noalign{\kern 10pt}
&&\hskip -1cm
\frac{d  \Sigma^{(0)}}{d \sp} (\sp=m) = 0 ~~,  \cr
\noalign{\kern 10pt}
&&\hskip -1cm
\Pi^{(0)} (\hp^2 =0) = 0 ~~, \cr
\noalign{\kern 10pt}
&&\hskip -1cm
\Gamma^{(0)M} ( \hp =\hp{\,}' ) = \gamma^M ~~.
\label{renormalization1}
\eeqn
One can also adopt the mass-independent renormalization where 
$m$ is set to be zero in the above equations (\ref{renormalization1}).
An alternative prescription is to impose the conditions on shell in 
$M^4 \times S^1$ for the 4D fermion with the lowest 4D mass
($\equiv m_{\rm phys}^{4D}$).  As is seen below, $V_\eff (\theta_H)$
has a global minimum at $\theta_H = \pi$ ($a=1/2 \equiv a_0 $).  Hence
the alternative renormalization conditions read
\beqn
&&\hskip -1cm
S(\hp) \Big|_{p^5={a_0 R^{-1}}} 
= \frac{i}{\sp - m - \Sigma} \bigg |_{p^5={a_0 R^{-1}}}
\sim \frac{i}{ \spp +a_0  R^{-1} \gamma^5  - m} \cr
\noalign{\kern 10pt}
&&\hskip  3.5cm
{\rm near}~~ p^2 = (m_{\rm phys}^{4D})^2 = m^2 +\frac{a_0^2}{R^2} ~ ,  \cr
\noalign{\kern 10pt}
&&\hskip -1cm
\Pi (p^2 =0, p^5=0) = 0 ~~, \cr
\noalign{\kern 10pt}
&&\hskip -1cm
\Gamma^{\mu} ( p =p{\,}' \, , p^5=p'{\,}^5=\frac{a_0}{R}) 
= \gamma^\mu ~~.
\label{renormalization2}
\eeqn
Renormalization with other reference values of $a_0$ is also possible.
In all of these prescriptions the Ward-Takahashi identity 
$Z_1=Z_2$ is preserved.

\section{Effective potential}

The effective potential $V_\eff$ for $\theta_H = 2\pi a$ is evaluated 
at the two loop level.
We adopt the dimensional regularization method to maintain the gauge
invariance.  The evaluation is performed in $M^d \times S^1$, and the 
$d \go 4$ limit is taken at the end.

$V_\eff (a)$ in $d+1$ dimensions at the one loop level is given by
\beeq
V_\eff (a)^{\rm 1 \, loop} = - \frac{f(d)}{2} \int \frac{d^d p_E}{(2\pi)^d}
\frac{1}{2\pi R} \sum_{n=-\infty}^\infty 
\ln \Big( p_E^2 + \frac{(n-a)^2}{R^2} + m^2 \Big) 
\label{effV1}
\eneq
after Wick rotation.  $f(d) = 2^{[(d+1)/2]}$.
In the $m \go 0$ limit it becomes
\beeq
V_\eff (a)^{\rm 1 \, loop} =  
\myfrac{f(d) \Gamma\Big(\mybig \frac{d+1}{2}\Big)}
  {(2 \pi^{3/2} R)^{d+1} } ~  f_{d+1}(a) 
+ {\rm constant}  ~~.  
\label{effV2}
\eneq
where $f_k(a)$ is defined in (\ref{function1}).  
The 4D effective potential is 
$V_\eff^{\rm 4D} (a) = V_\eff (a) \big|_{d=4} \times 2\pi R$ so that
\beeq
V_\eff^{\rm 4D} (a)^{\rm 1 \, loop} =  
\myfrac{3}{16 \pi^6 R^4} ~ f_5(a) 
+ {\rm constant}  ~~,
\label{effV3}
\eneq
where the constant is divergent, but is independent of $a$.

The effective potential is minimized at $a=1/2$ or $\theta_H = \pi$.
The effective potential for the zero mode of $A_5$, or the Higgs field $\phi_H$, 
is given by $V_\eff^{\rm 4D} (a)$ where $a$ is replaced by $e_4 R \phi_H$.  
Here the four-dimensional coupling is given by $e_4 = e/ \sqrt{2\pi R} \,$.
Hence the Higgs mass $m_H$ at the one loop level is given, for $m=0$,  by
\beeq
m_H^2 \big|_{\rm 1 \, loop} = e_4^2 R^2 \frac{d^2}{da^2} 
V_\eff^{\rm 4D} (a)^{\rm 1 \, loop} \Big|_{a=\onehalf}
= \myfrac{9 e_4^2 \, \zeta_R(3)}{16 \pi^4 R^2}  
\label{Higgs2}
\eneq
where $ \zeta_R(z)$ is the Riemann's zeta function.

\begin{figure}[tb]
\centering  \leavevmode
\includegraphics[width=10cm]{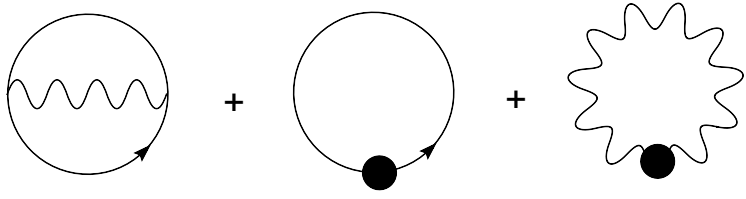}
\caption{Diagrams contributing to the effective potential
$V_\eff (\theta_H)$.  The second and third diagrams contain one loop
counter term $\delta_2 \sp - \delta_m$ and 
$\delta_3 (\hp^M \hp^N -  \eta^{MN} \hp^2)$, respectively.
}
\end{figure}

At the two loop level the diagrams in fig.\ 1  contribute to $V_\eff (a)$.
The contribution from the first diagram (a) is given by
\beqn
&&\hskip -1cm
- i V_\eff (a)^{\rm 2 \, loop \, (a)} 
= (-1) \frac{(-ie)^2}{2!} \int \frac{d^d p}{(2\pi)^d} \frac{1}{2\pi R} \sum_l
 \int \frac{d^d q}{(2\pi)^d} \frac{1}{2\pi R} \sum_n \cr
\noalign{\kern 10pt}
&&\hskip 2.cm
\times \frac{-i \eta_{MN}}{(\sp - \sq)^2 + i \ep} ~ 
\Tr \, \frac{i}{\sp - m + i\ep} \, \gamma^M \,
          \frac{i}{\sq - m + i\ep} \, \gamma^N   ~~, \cr
\noalign{\kern 10pt}
&&\hskip 3.cm
\hp^M = \Big(p^\mu, \frac{l-a}{R} \Big) ~~, ~~    
\hq^M = \Big(q^\mu, \frac{n-a}{R} \Big) ~~,
\label{2loopV1}
\eeqn
so that
\beqn
&&\hskip -1cm
 V_\eff (a)^{\rm 2 \, loop \, (a)} 
= - \frac{e^2}{2} \int \frac{d^d p_E}{(2\pi)^d}  \frac{d^d q_E}{(2\pi)^d}
\frac{1}{(2\pi R)^2} \sum_l  \sum_n   
\frac{f(d) \big\{ (d-1) \hp_E \hq_E + (d+1) m^2\big\} }
{(\hp_E^2 + m^2) \, ( \hq_E^2 + m^2) \,  (\hp_E - \hq_E)^2 } ~,
\cr
\noalign{\kern 10pt}
&&\hskip 0.cm
= - \frac{e^2 f(d)}{2} 
\int \frac{d^d p_E}{(2\pi)^d}  \frac{d^d q_E}{(2\pi)^d}
\frac{1}{(2\pi R)^2} \sum_l  \sum_n   \bigg[ 
\frac{2m^2}{(\hp_E^2 + m^2) \, ( \hq_E^2 + m^2) \,  (\hp_E - \hq_E)^2 } \cr
\noalign{\kern 10pt}
&&\hskip 2.cm
+ \frac{d-1}{2} \bigg\{ \frac{2}{(\hp_E^2 + m^2)  (\hp_E - \hq_E)^2}
- \frac{1}{(\hp_E^2 + m^2) \, ( \hq_E^2 + m^2)} \bigg\} \bigg] ~~.
\label{2loopV2}
\eeqn

At this stage infinite sums over discrete momentum $p_E^5$ have to be evaluated.
We summarize typical integral-sums in Appendix A.   In terms of $G_j(a; m, d)$
defined there, $V_\eff (a)^{\rm 2 \, loop \, (a)} $ can be expressed as
\beqn
&&\hskip -1cm
 V_\eff (a)^{\rm 2 \, loop \, (a)} 
=  - \frac{e^2 f(d)}{2}  \Big\{ 2m^2 G_2(a; m,d) \cr
\noalign{\kern 10pt}
&&\hskip 2.5cm
+ \frac{d-1}{2} \big[ 2 G_1(a; m,d) G_1 (0 ; 0,d) - G_1(a; m,d)^2 \big] \Big\} ~.
\label{2loopV3}
\eeqn
In the $mR \go 0$ limit (with $R$ kept fixed) it simplifies to
\beqn
&&\hskip -1cm
V_\eff (a)^{\rm 2 \, loop \, (a)} 
=  \frac{e^2(d-1)  f(d)}{4} \big\{  G_1(a; 0 ,d) -  G_1(0 ; 0 ,d) \big\}^2 \cr
\noalign{\kern 10pt}
&&\hskip .5cm
= \frac{e^2 (d-1) f(d) \Gamma \Big( \mybig \frac{d-1}{2} \Big)^2}
  {(4\pi)^{d+1}  (\pi R)^{2(d-1)} }
 \big\{  f_{d-1}(a) -  f_{d-1}(0) \big\}^2 
 \label{2loopV4}
\eeqn
up to an $a$-independent constant where $f_k(a)$ is defined in (\ref{function1}).
Its contribution to the 4D effective potential is given by
\beeq
V_\eff^{\rm 4D} (a)^{\rm 2 \, loop} =  
\myfrac{3e_4^2}{16 \pi^4 (2\pi R)^4} ~ \big\{  f_3(a) -  f_3(0) \big\}^2 ~.
\label{effV4}
\eneq
Contributions from one-loop counter terms, namely from the second and
third diagrams  in fig.\ 1,  either vanish for $m=0$ or are $a$-independent.

The effective potential at the two-loop level is given by
(\ref{effV3}) and (\ref{effV4}). The global minimum is located at $a=1/2$.
The two-loop contribution to the effective potential is
suppressed by an order of the fine structure constant, as seen 
in Eq. (\ref{effV4}). Hence, even though the two-loop
contribution itself is minimized at $a=0$, the effective
potential is governed by the one-loop 
contribution (\ref{effV3}) as long as the 
coupling $e_4^2/4\pi$ is small.  When one needs a very small value of
$a$ for having a  realistic model of the gauge-Higgs unification, the
two-loop contribution may play an important role and 
affect the location of the global
minimum of the effective potential as discussed in ref.\ \cite{Haba}.

The second derivative of   $V_\eff^{\rm 4D} (a)$ with respect to $a$ at the 
global minimum of $V_\eff^{\rm 4D}(a)$ is related to the coefficient 
$b_{-1} = - \Pi^{55} |_{p^5=0, p^2=0}$ introduced in Section 2, as easily
confirmed by examining Feynman diagrams.   Indeed,
\beeq
b_{-1} = e_4^2 R^2 \frac{d^2}{da^2} 
V_\eff^{\rm 4D} (a) \Big|_{a=\onehalf} ~~.
\label{Higgs3}
\eneq
Hence to the two loop order we have
\beeq
b_{-1} = \myfrac{9 e_4^2 \, \zeta_R(3)}{16 \pi^4 R^2}
 - \myfrac{21 e_4^4 \, \ln 2 \, \zeta_R(3)}{128 \pi^6 R^2} ~~.
 \label{Higgs4}
 \eneq
We would like to note that 
$e_4^2 R^2 (d^2V_\eff^{\rm 4D, \, 2 \, loop}/da^2) $ vanishes at $a=0$, 
which is in conformity  with the result  in ref.\ \cite{MY1}. 

\section{Vacuum polarization}

The vacuum polarization tensors $\Pi^{MN}$ in QED on $M^4 \times S^1$ 
has been evaluated to the two loop order near $a=0$ in Ref.\  \cite{MY1}.  
We need to determine  $\Pi^{MN}$ at $a= 1/2$ which corresponds to
the true vacuum.  In order to determine the Higgs mass (\ref{Higgs1}) 
to the two loop order, we need to find the coefficients $b_0$ and $c_0$,
or $\Pi^{MN}$ to the one loop order as $b_{-1}$ has been already evaluated
in (\ref{Higgs4}).   In this section  $\Pi^{MN} (\sp ; a, R)$ is evaluated
at the one loop level for an arbitrary value $a$.
We note that $\Pi^{MN}$ in supersymmetric gauge theory 
on an orbifold $M^4 \times (T^2/Z_2)$ with a vanishing Wilson line phase
has been evaluated at the one loop level.\cite{Ghilencea1}

The evaluation is straightforward.  The contribution from a fermion loop is
\beqn
&&\hskip -1cm
\Pi^{MN} (\hp) 
= i e^2 \int \frac{d^d q}{(2\pi)^d} \frac{1}{2\pi R} \sum_l 
\Tr \, \gamma^M \frac{1}{\sq - m + i\ep} \, \gamma^N \,
          \frac{1}{\sq + \sp - m + i\ep} \,   ~~, \cr
\noalign{\kern 10pt}
&&\hskip .5cm
\hp^M = \Big(p^\mu, \frac{n}{R} \Big) ~~, ~~    
\hq^M = \Big(q^\mu, \frac{l-a}{R} \Big) ~~.
\label{VacPol1}
\eeqn
In the dimensional regularization scheme the gauge invariance is
maintained so that  $\Pi^{MN}$ satisfies the current conservation;
\beeq
\hp_M \Pi^{MN} (\hp) = 0 ~~.
\label{current1}
\eneq

The detailed evaluation of $\Pi^{MN} (\hp) $ is given in Appendix B,
which is summarized in (\ref{VP3}) and (\ref{VP4}). 
To find the coefficients $b_0$ and $c_0$, we need $\Pi$ and $F$ 
at $p^5=n/R=0$.  From  (\ref{VP4}) and (\ref{VP5})  it follows that
\beqn
&&\hskip -1cm
\begin{pmatrix}
\Pi \cr F 
\end{pmatrix}_{p^5 = 0}
= 
\begin{pmatrix}
\Pi^{(0)} (p^2)  \cr 0 
\end{pmatrix}
- \frac{ e^2 f(d)}{(4\pi)^{(d+1)/2} } \int_0^1 dx \int_0^\infty dt \, t^{(1-d)/2}
e^{-t \{ m^2 - x(1-x) p^2 \} }  \cr
\noalign{\kern 15pt}
&&\hskip 4.0cm
\times
\sum_{\ell \not= 0}
e^{- \pi^2 R^2 \ell^2/t } e^{2\pi i \ell a}
\begin{pmatrix} 
2x(1-x)  \cr
\noalign{\kern 10pt}
\mybig \frac{1}{p^2} \frac{2\pi^2 \ell^2 R^2}{t^2}
\end{pmatrix} ~.  
\label{VacPol2}
\eeqn
The $\ell \not= 0$ terms give finite contributions.  At $d=4$ and $m=0$ 
\beqn
&&\hskip -1cm
\Pi |_{p^5=0} = \frac{3 e_4^2 R}{128} \, (-p^2)^{1/2} 
- \frac{e_4^2  f_1(a) }{6\pi^2}  + \cdots  ~~, \cr
\noalign{\kern 10pt}
&&\hskip -1cm
F |_{p^5=0} = - \frac{3 e_4^2 f_3(a)}{4\pi^4 R^2}  \, \frac{1}{p^2}
-\frac{e_4^2  f_1(a) }{12 \pi^2}  + \cdots   ~~.
\label{VacPol3}
\eeqn
Note that the counter term $\delta_3 =  \Pi^{(0)} (\hp^2=0) $ vanishes at $m=0$.
We expand the invariant functions around $p^2=0$  at the global minimum 
of the $V_\eff(a)$, namely at $a=1/2$;
\beqn
&&\hskip -1cm
\Pi |_{p^5=0 , \, a=\onehalf} = c_0 + \cdots ~~, \cr
\noalign{\kern 10pt}
&&\hskip -1cm
F |_{p^5=0 , \, a=\onehalf} = \frac{b_{-1}}{p^2} + b_0 + \cdots ~~.
\label{coeff1}
\eeqn
The coefficients at the one loop level are given by
\beqn
&&\hskip -1cm
c_0 = \frac{e_4^2 \,  \ln 2 }{6\pi^2}  ~~, \cr
\noalign{\kern 10pt}
&&\hskip -1cm
b_0 = \frac{e_4^2 \,  \ln 2 }{12 \pi^2} ~~, \cr
\noalign{\kern 10pt}
&&\hskip -1cm
b_{-1} =  \frac{9 \, e_4^2 \, \zeta_R(3)}{16 \pi^4 R^2}  ~~.
\label{coeff2}
\eeqn
The coefficient $b_{-1}$ coincides with the result from the effective potential
(\ref{Higgs2}) or (\ref{Higgs4}) as it should.

The gauge invariant mass for the zero mode of $A_5$ appears as a pole
$1/p^2$ in $F$.   We remark that there is similarity to the gauge invariant
mass in the Schwinger model (QED in two dimensions) in which $\Pi$ 
develops a pole from a fermion loop.\cite{Schwinger}
It differs in the point that $F$ vanishes
in the $R \go \infty$ limit, or in $M^5$, whereas the pole remains 
in the Schwinger model in $M^2$.  The effective potential for $\theta_H$ in the 
Schwinger model on a circle has the same structure as in the current
model.\cite{HH}  Its curvature at the minimum gives a mass for photons.

\section{The Higgs mass}

The 4D effective action for the Higgs field $\phi_H$ takes  the form
\beeq
\Gamma_\eff [ \phi_H] = \int d^4 x \, \Big\{
- V[\phi_H] 
 + \onehalf Z[\phi_H] \dd_\mu \phi_H \dd^\mu \phi_H
+ \cdots \Big\} ~~.
\label{effAction1}
\eneq
The Higgs mass in this approach is 
\beeq
m_H^2 = \myfrac{1}{Z[\phi_H]} 
\myfrac{\dd^2 V[\phi_H]}{\dd \phi_H^2} \Bigg|_{\phi_H^{\rm min}}
\label{Higgs6}
\eneq
where $\phi_H^{\rm min}= (2 e_4 R)^{-1}$ is the location of the global minimum of 
$V[\phi_H]$.   The effective potential $V[\phi_H]$ is given by
$V_\eff^{\rm 4D} (a)$ with $a= e_4 R\phi_H$.  Its second derivative is 
related to $b_{-1}$ by (\ref{Higgs3}).  
Similarly $Z[\phi_H^{\rm min}]$ is related to $c_0$ and $b_0$ by
$Z[\phi_H^{\rm min}] = 1 - c_0 - b_0$.  Hence the Higgs mass defined 
by (\ref{Higgs6}) coincides with the mass defined by  (\ref{Higgs1}).
Inserting (\ref{Higgs4}) and $c_0, b_0$ in (\ref{coeff2}) there, 
one finds that 
in the massless fermion limit $m=0$
\beeq
m_H^2 = \myfrac{9 \, e_4^2 \, \zeta_R(3)}{16 \pi^4 R^2}
\bigg\{ 1 - \myfrac{e_4^2 \, \ln 2}{24 \pi^2} \bigg\} ~~.
\label{Higgs5}
\eneq

The coupling constant $e$ and the coefficients $b_{-1}, b_0, c_0$  
depend on the renormalization scheme.   
However, the Higgs mass is a physical quantity so that it should not depend on the
renormalization scheme employed.   This can be confirmed from the results at the 
two loop level obtained above.

Let $e'$, $b_{-1}', b_0', c_0'$ be the coupling constant and 
the coefficients in  a second renormalization scheme.  To be concrete,
$(e, b_{-1}, b_0, c_0)$ are defined in the renormalization in $M^5$
as employed in the preceding sections, whereas   
$(e', b_{-1}', b_0', c_0')$ are defined in the on-shell renormalization 
in $M^4 \times S^1$ at $a=1/2$.  For the sake of simplicity
we suppose that $m=0$.  At the one loop level we write
$\Pi(\hp) = \Pi_r (\hp) - \delta_3$ where $\Pi_r (\hp)$ is the 
contribution from a fermion loop.   The counter terms are given by
\beqn
&&\hskip -1cm
\delta_3 = \lim_{R\go \infty} \Pi_r (\hp \, ; a, R) \big|_{\hp^2 = 0} ~~, \cr
\noalign{\kern 10pt}
&&\hskip -1cm
\delta_3' =  \Pi_r (p^2=0, p^5=0  ; a=\onehalf , R) ~~. 
\label{renormal1}
\eeqn
The difference between the two, $\delta_3 - \delta_3'$,  is finite. 
The coefficient $c_0$ is defined
by the expansion  of $\Pi (p^2, p^5=0) = c_0 + \cdots$ in $p^2$. 
Hence $c_0' - c_0 = \delta_3 - \delta_3'$.  The Ward-Takahashi 
identity $Z_1=Z_2$ implies  $e^{(0)} = Z_3^{-1/2} e$.   It follows that
\beeq
e'^2 = \frac{Z_3'}{Z_3} ~  e^2  \simeq \frac{e^2}{1 - c_0 + c_0'} ~~.
\label{renormal2}
\eneq

We observed that $F$ is finite at the one loop level in Section 4 and
$V_\eff (a)^{\rm 2 \, loop \, (a)}$ itself is finite in the $mR \go 0$ limit
in Section 3.    We write 
$b_{-1} = e^2 {b_{-1}}^{(1)} +  e^4 {b_{-1}}^{(2)} + \cdots$ and
$b_0 = e^2 {b_0}^{(1)}  + \cdots$.  
Then the finiteness implies that
${b_{-1}}^{(1)} = {b'_{-1}}^{(1)}$, 
${b_{-1}}^{(2)} = {b'_{-1}}^{(2)}$,  and
${b_{0}}^{(1)} = {b'_{0}}^{(1)}$.  From these identities one finds that
\beqn
&&\hskip -1cm
(m_H^2)' = \frac{b_{-1}'}{1 - c_0' - b_0'} 
= e'^2 \cdot \frac{ {b'_{-1}}^{(1)} + e'^2 {b'_{-1}}^{(2)} }
{1 - c_0' - e'^2 {b'_0}^{(1)}} \cr
\noalign{\kern 10pt}
&&\hskip .2cm
\simeq \frac{e^2}{1-c_0 + c_0'} \cdot 
\frac{ {b_{-1}}^{(1)} + e^2 {b_{-1}}^{(2)} }
{1 - c_0' - e^2 {b_0}^{(1)}} \cr
\noalign{\kern 10pt}
&&\hskip .2cm
\simeq  \frac{b_{-1}}{1 - c_0 - b_0} = m_H^2 ~~.
\label{Higgs7}
\eeqn
The Higgs mass is independent of the renormalization scheme to
this order as it should be.

\section{Summary and discussions}

In this paper we have determined the Higgs mass $m_H$ at the two loop level,
or to $O(e^4)$, 
in the QED gauge-Higgs unification model on $M^4 \times S^1$.   
The mass is shown to be  independent of the renormalization scheme. 
The evaluation of the vacuum polarization tensors, or equivalently
$Z[\phi_H]$ in the effective action, at the one loop level
is also required to find $m_H$ at the two loop level.  
The $\theta_H$-dependent part of the effective potential is found  
finite at the two loop level.  Divergences in $\Pi^{MN} (\hp)$, which  appear
only in the $\Pi$ part, but not in the $F$ part, 
are absorbed by the counter term $\delta_3$.  There is no need to introduce
additional counter terms other than $\delta_1$, $\delta_2$, $\delta_3$ and 
$\delta_m$ at this level.

The fact that radiative corrections to the Higgs mass are finite and 
suppressed by a power of the four-dimensional gauge coupling constant 
with respect to the Kaluza-Klein mass scale $m_{KK} = 1/R$ has an 
important implication in the gauge-Higgs unification.  
The Higgs mass is not an input parameter of the theory, but is definitively
predicted in terms of other fundamental constants such as the gauge coupling
and the size of the extra dimension.  Its value is stable
against higher order corrections.  In other words the gauge-Higgs unification yields a 
naturally light Higgs boson in four dimensions.  
We remark that in the gauge-Higgs unification in flat space, however, 
the Higgs mass becomes small compared with the $W$ and $Z$ boson masses 
unless the Wilson line phase $\theta_H$ is sufficiently 
small.\cite{HNT2, Haba, Csaki2, Sakamoto1}
This problem can be naturally resolved in the gauge-Higgs unification in the 
warped space.\cite{HM}       Further the Higgs interactions 
with other fields and particles can be predicted 
as well.\cite{HNSS, SH1, HS2, Sakamura, Falkowski2}

Although we considered, for the sake of simplicity, 
 the massless fermion limit ($mR \go 0$) to find $m_H$
in the present paper, the same 
features are expected to hold in the $m \not= 0$ case.   Contributions of
non-vanishing $\delta_1$, $\delta_2$ and $\delta_m$ have to be taken 
into account.  In passing, we would like to point out that the massless 
fermion limit is well defined on $M^4 \times S^1$ and on $M^3 \times S^1$.
As the effective potential is minimized at $a= 1/2$, the fermion 
propagator does not vanish at $\hp^M=0$ with $a= 1/2$.

Extension of our analysis to QED in $M^5 \times S^1$ is straightforward.
To define a theory and determine a Higgs mass at the two loop level, 
one must include additional counter terms such as $(\dd_L F_{MN})^2$
in the original Lagrangian.  Other than this the analysis remains intact with 
the substitution $d=5$.

More important is the extension to the non-Abelian case and to higher
order corrections in the viewpoint of the gauge-Higgs unification. 
Not only propagators of gauge fields have $\theta_H$ dependence, 
but also there appear a new interaction vertex proportional to $\theta_H$.
It is curious to see how the large gauge invariance
($\theta_H \go \theta_H + 2\pi$) is maintained in perturbation theory.

\vskip .5cm

\leftline{\large \bf Acknowledgments}
This work was supported in part by  Scientific Grants from the Ministry of 
Education and Science, Grant No.\ 17540257(Y.H.), 
Grant No.\ 18204024(Y.H. and N.M.),  
and Grant No.\ 19034007(Y.H.).    K.T. is supported by the 21st Century COE 
Program at Tohoku University.  One of the authors (Y.H.) would 
like to thank the CERN Theory Institute for its hospitality where
a part of this work was done. 

\vskip .5cm

\appendix

\section{Integrals and sums}

We summarize  useful formulas for the evaluation in Section 3.
The first integral-sum is
\beqn
&&\hskip -1cm
G_1(a ; m, d) = 
\int \frac{d^d p_E}{(2\pi)^d}   \frac{1}{2\pi R} 
\sum_{\ell = -\infty}^\infty
      \frac{1}{\hp_E^2 + m^2}   \qquad (p_E^5 = \frac{\ell -a}{R} )\cr
\noalign{\kern 10pt}
&&\hskip 1cm
= \int_0^\infty dt \, \frac{e^{-tm^2}}{(4\pi t)^{d/2}}
\frac{1}{2\pi R} \sum_{\ell =-\infty}^\infty    e^{-t(\ell -a)^2/R^2}  \cr
\noalign{\kern 10pt}
&&\hskip 1cm
=  \int_0^\infty dt \, \frac{e^{-tm^2}}{(4\pi t)^{(d+1)/2}} 
\sum_{\ell =-\infty}^\infty  e^{-\pi^2 R^2 \ell^2/t} ~  e^{-2\pi i \ell a} \cr
\noalign{\kern 10pt}
&&\hskip 1cm
= \frac{\Gamma \Big( \mybig \frac{1-d}{2} \Big)  m^{d-1}}{(4\pi)^{(d+1)/2}}
+ \frac{m^{d-1} }{(2\pi)^d}  \sum_{\ell =1}^\infty 
    \frac{2 \cos (2\pi \ell a)   K_{(d-1)/2} (2\pi\ell mR)}{(\ell mR)^{(d-1)/2} }  ~~.
\label{sum1}
\eeqn
In the third equality the Poisson resummation formula has been employed.
In the last expression $K_\nu(z)$ is the modified Bessel function.
For small $mR$ one finds
\beqn
&&\hskip -1cm
G_1(a ; m, d) =  
\frac{\Gamma \Big( \mybig \frac{1-d}{2} \Big)  m^{d-1}}{(4\pi)^{(d+1)/2}}
+\frac{2\Gamma \Big( \mybig \frac{d-1}{2} \Big)  f_{d-1}(a)}{(4\pi)^{(d+1)/2} (\pi R)^{d-1}}
- \frac{2\Gamma \Big( \mybig \frac{d-3}{2} \Big)  f_{d-3}(a) m^2}
{(4\pi)^{(d+1)/2} (\pi R)^{d-3}}
+ \cdots
\label{sum2}
\eeqn
where
\beeq
f_k(a) =  \sum_{\ell =1}^\infty  \frac{\cos 2\pi \ell a}{\ell^k} ~~.
\label{function1}
\eneq

The second  integral-sum is
\beqn
&&\hskip -1cm
G_2(a ; m, d) = 
\int \frac{d^d p_E}{(2\pi)^d}  \frac{d^d q_E}{(2\pi)^d}
\frac{1}{(2\pi R)^2} \sum_{\ell =-\infty}^\infty   \sum_{n=-\infty}^\infty    
\frac{1}{(\hp_E^2 + m^2) \, ( \hq_E^2 + m^2) \,  (\hp_E - \hq_E)^2 } \cr
\noalign{\kern 10pt}
&&\hskip 3cm
\Big( ~ p_E^5 = \frac{\ell -a}{R} ~,~ q_E^5 = \frac{n -a}{R}  ~ \Big) ~.
\label{sum3}
\eeqn
Introducing Feynman parameters,  exponentiating the denominator,  
integrating over $p_E$ and $q_E$,  and making repeated use of
the Poisson resummation formula, one finds
\beqn
&&\hskip -1cm
G_2(a ; m, d) = \frac{1}{2} \int_\Omega dxdy    \int_0^\infty dt \,
\frac{t^2 e^{- t(x+y) m^2}}{(4\pi t)^{d+1} h(x,y)^{(d+1)/2}}    \cr
\noalign{\kern 10pt}
&&\hskip 1.5cm
\times  \sum_{\ell =-\infty}^\infty   \sum_{n=-\infty}^\infty  e^{-2\pi i (\ell +n)a}
\exp \bigg\{ - \frac{ \pi^2 R^2 S_{\ell n} (x,y) }{ t~ h(x,y)} \bigg\}~,\cr
\noalign{\kern 15pt}
&&\hskip -.5cm
\Omega = \{ (x, y) ; 0 \le x, y, x+y \le 1 \} ~, \cr
\noalign{\kern 5pt}
&&\hskip -.5cm
h(x,y) = (1-x)(1-y) -(1-x-y)^2  ~, \cr
\noalign{\kern 5pt}
&&\hskip -.5cm
S_{\ell n} (x,y) = (1-x) \ell^2 + (1-y) n^2 + 2(1-x-y) \ell n ~~.
\label{sum4}
\eeqn
The $(\ell , n) = (0,0)$ term gives contributions in $M^5$, which is independent of $a$.
For small $mR$ one finds
\beqn
&&\hskip -1cm
G_2(a ; m, d) =\frac{ \Gamma(2-d) \, m^{2(d-2)} }{ 2 (4\pi)^{d+1} } 
\int_\Omega dxdy \, \frac{(x+y)^{d-2}}{h(x,y)^{(d+1)/2} }  \cr
\noalign{\kern 10pt}
&&\hskip 0.cm
+ \frac{ \Gamma(d-2)}{ 2 (4\pi)^{d+1}  (\pi R)^{2(d-2)} } 
\sum_{(\ell,n) \not= (0,0)}  e^{-2\pi i (\ell +n)a}
\int_\Omega dxdy \, \frac{h(x,y)^{(d-5)/2} }{ S_{\ell n} (x,y)^{d-2}}\cr
\noalign{\kern 10pt}
&&\hskip 0.cm
- \frac{ \Gamma(d-3) \, m^2}{ 2 (4\pi)^{d+1}  (\pi R)^{2(d-3)} } 
\sum_{(\ell,n) \not= (0,0)}  e^{-2\pi i (\ell +n)a}
\int_\Omega dxdy \, \frac{h(x,y)^{(d-7)/2} }{ S_{\ell n} (x,y)^{d-3}} ~  + \cdots ~. 
\label{sum5}
\eeqn

\section{Evaluation of $\Pi^{MN}$}

Evaluation of the vacuum polarization tensors proceeds as follows.
Performing the trace in (\ref{VacPol1}) and introducing a Feyman parameter, 
one obtains
\beqn
&&\hskip -1cm
\Pi^{MN} (\hp) 
=  i e^2 f(d) \int_0^1 dx \int \frac{d^d q}{(2\pi)^d} \frac{1}{2\pi R} \sum_l
\frac{S^{MN}}{[ \hq^2 - m^2 + x(\hp^2 + 2 \hq \hp)  + i\ep  ]^2} ~, \cr
\noalign{\kern 10pt}
&&\hskip .5cm
S^{MN} = 2 \hq^M \hq^N + \hq^M \hp^N + \hp^M \hq^N 
  - \hq (\hq + \hp) \eta^{MN} + m^2 \eta^{MN} ~~.
\label{VP1}
\eeqn
We shift the integration variable $q^\mu \go q'^\mu  = q^\mu + x p^\mu$, 
exponentiate the denominator, and integrate over Wick-rotated $q'_E$
to find
\beqn
&&\hskip -1cm
\begin{pmatrix}
\Pi^{\mu\nu} \cr \Pi^{55} \cr \Pi^{\mu 5} 
\end{pmatrix}
= - \frac{e^2 f(d)}{(4\pi)^{d/2}}  \int_0^1 dx \int_0^\infty dt \, t^{1-(d/2)}
\frac{1}{2\pi R} \sum_{\ell = -\infty}^\infty
e^{-t [ (q^5 + xp^5)^2 + m^2 - x(1-x) \hp^2 ]} \cr
\noalign{\kern 10pt}
&&\hskip 0.cm
\times
\begin{pmatrix}
\big\{  (\onehalf  d -1) t^{-1} + x(1-x) p^2 + q^5 (q^5+p^5) + m^2 \big\} \eta^{\mu\nu}
    - 2x(1-x) p^\mu p^\nu \cr
    - \onehalf t^{-1} - x(1-x) p^2 + q^5 (q^5+p^5)  - m^2 \cr
    p^\mu \big\{ (1-2x) q^5 - x p^5 \big\} 
\end{pmatrix} ~.
\label{VP2}
\eeqn
Recalling $q^5 = (\ell -a)/R$ and $p^5 = n/R$, one employ the Poisson resummation 
formula to find
\beqn
&&\hskip -1cm
\begin{pmatrix}
\Pi^{\mu\nu} \cr \Pi^{55} \cr \Pi^{\mu 5} 
\end{pmatrix}
= \frac{- e^2 f(d)}{(4\pi)^{(d+1)/2} } \int_0^1 dx \int_0^\infty dt \, t^{(1-d)/2}
e^{-t \{ m^2 - x(1-x) \hp^2 \} }
\sum_{\ell = -\infty}^\infty
e^{- \pi^2 R^2 \ell^2/t } e^{2\pi i \ell (a- xn)}  \cr
\noalign{\kern 10pt}
&&\hskip .0cm
\times
\begin{pmatrix}
\Big\{ \mybig \frac{d-1}{ 2t}  - \frac{\pi^2 R^2 \ell^2}{ t^2} +(1-2x) \frac{ i\pi n\ell}{t} 
    + x(1-x) \hp^2 + m^2 \Big\} \eta^{\mu\nu}     \cr
\hskip 9cm     - 2x(1-x) p^\mu p^\nu  \cr
\noalign{\kern 8pt}
- \mybig \frac{d-1}{2t} - \frac{\pi^2 R^2 \ell^2}{t^2} + (1-2x)  \frac{i\pi n\ell}{t} 
  - x(1-x)  \big\{ (p^5)^2 + p^2  \big\} - m^2 \cr
\noalign{\kern 8pt}
p^\mu \Big\{ \mybig  (1-2x) \frac{ i\pi R \ell}{t}  - 2x(1-x) p^5 \Big\}
\end{pmatrix} ~.
\label{VP2-1}
\eeqn

The expression is simplified with  identities 
\beqn
&&\hskip -1cm
 \int_0^\infty dt \,    \Big\{ \frac{1-d}{2t} - m^2  + x(1-x) \hp^2 
    + \frac{\pi^2 R^2 \ell^2}{ t^2}    \Big\} 
t^{(1-d)/2} e^{-t \{ m^2 - x(1-x) \hp^2 \} }e^{- \pi^2 R^2 \ell^2/t }  \cr
\noalign{\kern 10pt}
&&\hskip 3.0cm
=  \int_0^\infty dt \, \frac{\dd}{\dd t} \Big\{ 
t^{(1-d)/2} e^{-t \{ m^2 - x(1-x) \hp^2 \} }
e^{- \pi^2 R^2 \ell^2/t }   \Big\} = 0 ~, \cr
\noalign{\kern 10pt}
&&\hskip -1.0cm
\int_0^1 dx \Big\{ (1-2x) \hp^2 - \frac{2\pi i\ell n}{t} \Big\} 
e^{tx(1-x) \hp^2 - 2\pi i \ell n x}
= \int_0^1 dx \, \frac{1}{t} \frac{\dd}{\dd x} 
e^{tx(1-x) \hp^2 - 2\pi i \ell n x}  = 0 ~,
\label{identity1}
\eeqn
to 
\beqn
&&\hskip -1cm
\begin{pmatrix}
\Pi^{\mu\nu} \cr \Pi^{55} \cr \Pi^{\mu 5} 
\end{pmatrix}
= \frac{- e^2 f(d)}{(4\pi)^{(d+1)/2} } \int_0^1 dx \int_0^\infty dt \, t^{(1-d)/2}
e^{-t \{ m^2 - x(1-x) \hp^2 \} }
\sum_{\ell = -\infty}^\infty
e^{- \pi^2 R^2 \ell^2/t } e^{2\pi i \ell (a- xn)}  \cr
\noalign{\kern 10pt}
&&\hskip -.5cm
\times
\left\{
2x(1-x) \big( \hp^2 \eta^{MN} - \hp^M \hp^N \big) 
+ (1-2x) \frac{i\pi \ell}{t} 
\begin{pmatrix}
n \eta^{\mu\nu} \cr n \cr  p^\mu R 
\end{pmatrix}
- \frac{2\pi^2 R^2 \ell^2 }{ t^2}
\begin{pmatrix}
0 \cr 1  \cr  0 
\end{pmatrix}
\right\}   ~.
\label{VP3}
\eeqn
Notice that only the $\ell=0$ term survives in the $R \go \infty$ limit. 
$\Pi^{\mu\nu}$ and $\Pi^{55}$ are even in $p^5=n/R$, whereas
$\Pi^{\mu 5}$ is odd after the integration over $x$.

As a consequence of the current conservation $\Pi^{MN}$ can be
expressed in terms of the two invariant functions $\Pi$ and $F$ in 
(\ref{invariant2}).  Comparison of the two expressions for $\Pi^{55}$ and
$\Pi^{\mu 5}$ in (\ref{VP3}) shows that in the integrand 
$- 2\pi^2 R^2 \ell^2/t^2$ and $(\hp^2/p^5) (1-2x) R (i\pi\ell/t)$ give
the same contribution in  (\ref{VP3}).  

The invariant functions are given by
\beqn
&&\hskip -1cm
\begin{pmatrix}
\Pi  \cr F
\end{pmatrix}
= \frac{- e^2 f(d)}{(4\pi)^{(d+1)/2} } \int_0^1 dx \int_0^\infty dt \, t^{(1-d)/2}
e^{-t \{ m^2 - x(1-x) \hp^2 \} }  \cr
\noalign{\kern 7pt}
&&\hskip .5cm
\times
\sum_{\ell = -\infty}^\infty
e^{- \pi^2 R^2 \ell^2/t } e^{2\pi i \ell (a- xn)}
\begin{pmatrix} \mybig
2x(1-x) + \frac{1}{\hp^2} (1-2x) n \frac{i\pi \ell}{t} \cr
\noalign{\kern 10pt}
\mybig \frac{p^2}{(\hp^2)^2} \frac{2\pi^2 \ell^2 R^2}{t^2}
\end{pmatrix} ~.  
\label{VP4}
\eeqn
In the $R \go \infty$ limit, namely in $M^{d+1}$, $F$ vanishes and
\beqn
&&\hskip -1cm 
\Pi^{(0)} (\hp^2) = \lim_{R \go \infty} \Pi  \cr
\noalign{\kern 10pt}
&&\hskip 0.5cm
= - \frac{2 e^2 f(d) \Gamma \Big( \mybig\frac{3-d}{2} \Big) }
            {(4\pi)^{(d+1)/2} }
\int_0^1 dx \, x(1-x) \big[ m^2 - x(1-x) \hp^2 \big]^{(d-3)/2} ~.
\label{VP5}
\eeqn
At $d=4$ ($D=5$),  $\Pi^{(0)} (0)$ is removed by the counter term $\delta_3$
so that the renormalized $\Pi$ is $\Pi (\hp) - \Pi^{(0)} (0)$.
At $d=5$ ($D=6$) an additional counter term $(\dd_L F_{MN})^2$
is necessary to remove the  divergence proportional to $\hp^2$ in 
(\ref{VP5}).

\def\jnl#1#2#3#4{{#1}{\bf #2} (#4) #3}

\def\Zphys{{\em Z.\ Phys.} }
\def\jssc{{\em J.\ Solid State Chem.\ }}
\def\jpsJ{{\em J.\ Phys.\ Soc.\ Japan }}
\def\ptps{{\em Prog.\ Theoret.\ Phys.\ Suppl.\ }}
\def\PTP{{\em Prog.\ Theoret.\ Phys.\  }}

\def\JMP{{\em J. Math.\ Phys.} }
\def\NPB{{\em Nucl.\ Phys.} B}
\def\NP{{\em Nucl.\ Phys.} }
\def\PLB{{\em Phys.\ Lett.} B}
\def\PL{{\em Phys.\ Lett.} }
\def\PRL{\em Phys.\ Rev.\ Lett. }
\def\PRB{{\em Phys.\ Rev.} B}
\def\PRD{{\em Phys.\ Rev.} D}
\def\PRe{{\em Phys.\ Rep.} }
\def\AP{{\em Ann.\ Phys.\ (N.Y.)} }
\def\RMP{{\em Rev.\ Mod.\ Phys.} }
\def\ZPC{{\em Z.\ Phys.} C}
\def\SCI{\em Science}
\def\CMP{\em Comm.\ Math.\ Phys. }
\def\MPLA{{\em Mod.\ Phys.\ Lett.} A}
\def\IJMPA{{\em Int.\ J.\ Mod.\ Phys.} A}
\def\IJMPB{{\em Int.\ J.\ Mod.\ Phys.} B}
\def\EPJC{{\em Eur.\ Phys.\ J.} C}
\def\PR{{\em Phys.\ Rev.} }
\def\JHEP{{\em JHEP} }
\def\cmp{{\em Com.\ Math.\ Phys.}}
\def\JPA{{\em J.\  Phys.} A}
\def\JPG{{\em J.\  Phys.} G}
\def\NJP{{\em New.\ J.\  Phys.} }
\def\CQG{\em Class.\ Quant.\ Grav. }
\def\ATMP{{\em Adv.\ Theoret.\ Math.\ Phys.} }
\def\ibid{{\em ibid.} }

\renewenvironment{thebibliography}[1]
         {\begin{list}{[$\,$\arabic{enumi}$\,$]}  
         {\usecounter{enumi}\setlength{\parsep}{0pt}
          \setlength{\itemsep}{0pt}  \renewcommand{\baselinestretch}{1.2}
          \settowidth
         {\labelwidth}{#1 ~ ~}\sloppy}}{\end{list}}

\def\reftitle#1{}                


\vskip .8cm

\begin{thebibliography}{99}
\small
\baselineskip=14pt


\leftline{\large \bf References}




\bibitem{Fairlie1}
D.B.\ Fairlie, \jnl{\PLB}{82}{97}{1979};
\reftitle{Higgs' Fields And The Determination of The Weinberg Angle}
\jnl{\JPG}{5}{L55}{1979}.
\reftitle{Two Consistent Calculations of The Weinberg Angle}

\bibitem{Manton1}
N.\ Manton, \jnl{\NPB}{158}{141}{1979};
\reftitle{A New Six-Dimensional Approach to the Weinberg-Salam Model}

P.\ Forgacs and N.\ Manton, \jnl{\CMP}{72}{15}{1980}.
\reftitle{Space-Time Symmetries In Gauge Theories}



\bibitem{YH1}
Y.\ Hosotani, \jnl{\PLB}{126}{309}{1983}.
\reftitle{Dynamical Mass Generation By Compact Extra Dimensions}

\bibitem{YH4}
Y.\ Hosotani,  \jnl{\PLB}{129}{193}{1984}; 
\reftitle{Dynamical Gauge Symmetry Breaking As The Casimir Effect}
\jnl{\PRD}{29}{731}{1984}.
\reftitle{Dynamical Gauge Symmetry Breaking and Left-Right Asymmetry In Higher Dimensional Theories}



\bibitem{YH2}
Y.\ Hosotani, \jnl{\AP}{190}{233}{1989}.
\reftitle{Dynamics of nonintegrable phases and gauge symmetry breaking}

\bibitem{McLachlan1}
A.T.\ Davies and A.\ McLachlan,
\jnl{\PLB}{200}{205}{1988};
\reftitle{Gauge group breaking by Wilson loops}
\jnl{\NPB}{317}{237}{1989}.
\reftitle{Congruency class effects in the Hosotani model}

\bibitem{McLachlan2}
A.\ McLachlan,
\jnl{\PLB}{222}{372}{1989};
\reftitle{Flux breaking of $E_6$ in the generalized Hosotani model}
\jnl{\NPB}{338}{188}{1990}.
\reftitle{Flux-breaking in space-times with toroidal compactification}

\bibitem{Takenaga1}
K.\ Takenaga, 
\jnl{\PLB}{425}{114}{1998};
\reftitle{Supersymmetry breaking through boundary conditions associated with the U(1)(R)}
\jnl{\PRD}{58}{026004}{1998}, 
{\it Erratum}, \jnl{\ibid}{{\rm D}61}{129902}{2000}.
\reftitle{Softly broken supersymmetric gauge theories through compactifications}

\bibitem{Lim1}
H.\ Hatanaka, T.\ Inami and C.S.\ Lim, 
\jnl{\MPLA}{13}{2601}{1998}.
\reftitle{The Gauge Hierarchy Problem and Higher Dimensional Gauge Theories}




\bibitem{Pomarol1}
A.\ Pomarol and M.\ Quiros, \jnl{\PLB}{438}{255}{1998};
\reftitle{The Standard Model from extra dimensions}


\bibitem{Antoniadis1}
I.\ Antoniadis, K.\ Benakli and M.\ Quiros,
\jnl{\it New. J.\ Phys.}{3}{20}{2001}.
\reftitle{Finite Higgs mass without Supersymmetry}

\bibitem{Csaki1}
C.\ Csaki, C.\ Grojean and H.\ Murayama, \jnl{\PRD}{67}{085012}{2003};
\reftitle{Standard Model Higgs From Higher Dimensional Gauge Fields}

C.A.\ Scrucca, M.\ Serone and L.\ Silverstrini, \jnl{\NPB}{669}{128}{2003}. 
\reftitle{Electroweak symmetry breaking and fermion masses from extra dimensions}

\bibitem{gaugeHiggs3}
L.J.\ Hall, Y.\ Nomura and D.\ Smith,  \jnl{\NPB}{639}{307}{2002};
\reftitle{Gauge-Higgs Unification in Higher Dimensions}

L.\ Hall, H.\ Murayama, and Y.\ Nomura, 
   \jnl{\NPB}{645}{85}{2002};
\reftitle{Wilson Lines and Symmetry Breaking on Orbifolds}

G.\ Burdman and Y.\ Nomura, \jnl{\NPB}{656}{3}{2003}; 
\reftitle{Unification of Higgs and Gauge Fields in Five Dimensions}



C.A.\ Scrucca, M.\ Serone, L.\ Silvestrini and A.\ Wulzer,
\jnl{\JHEP}{0402}{49}{2004};
\reftitle{Gauge-Higgs Unification in Orbifold Models} 

\bibitem{HHHK}
N.\ Haba, M.\ Harada, Y.\ Hosotani and Y.\ Kawamura, 
\jnl{\NPB}{657}{169}{2003};   
{\it Erratum}, {\it ibid.}  B{\bf 669} (2003) {381}.
\reftitle{Dynamical Rearrangement of Gauge Symmetry on the Orbifold $S^1/Z_2$}



\bibitem{HHKY}
N.\ Haba,  Y.\ Hosotani,  Y.\ Kawamura and T.\ Yamashita, 
\jnl{\PRD}{70}{015010}{2004};
\reftitle{Dynamical symmetry breaking in Gauge-Higgs unification on orbifold}


N.\ Haba and T.\ Yamashita, \jnl{\JHEP}{0404}{016}{2004}. 
\reftitle{Dynamical symmetry breaking in gauge Higgs unification 
of 5D N=1 SUSY theory} 

\bibitem{HY}
N.\ Haba and T.\ Yamashita, \jnl{\JHEP}{0402}{059}{2004}. 
\reftitle{A General formula of the effective potential in 5-D SU(N) gauge theory on orbifold}



\bibitem{Lim3}
K.\ Hasegawa, C.S.\ Lim and N.\ Maru, \jnl{\PLB}{604}{133}{2004}. 
\reftitle{An Attempt to solve the hierarchy problem based on 
gravity-gauge-Higgs unification scenario}

\bibitem{HNT2}
Y.\ Hosotani, S.\ Noda and K.\ Takenaga,
\jnl{\PLB}{607}{276}{2005}.
\reftitle{Dynamical Gauge-Higgs Unification in the Electroweak Theory}

\bibitem{Haba}
N.\ Haba,  K.\ Takenaga, and T.\ Yamashita, 
\jnl{\PLB}{615}{247}{2005}.
\reftitle{Higgs mass in the gauge-Higgs unification}




\bibitem{Csaki2}
G.\ Cacciapaglia, C.\ Csaki and S.C.\ Park,
\jnl{\JHEP}{0603}{099}{2006}.
\reftitle{Fully Radiative Electroweak Symmetry Breaking}

\bibitem{Panico2}
G.\ Panico, M.\ Serone and A.\ Wulzer,
\jnl{\NPB}{739}{186}{2006}.
\reftitle{A Model of Electroweak Symmetry Breaking from a Fifth Dimension}


\bibitem{DiazCruz}
A.\ Aranda and J.L.\ Diaz-Cruz,  
\jnl{\PLB}{633}{591}{2006}.
\reftitle{Gauge-Higgs unification with brane kinetic terms}

\bibitem{Sakamoto1}
M.\ Sakamoto and K.\ Takenaga, \jnl{\PRD}{75}{045015}{2007}. 
\reftitle{Large gauge hierarchy in gauge-Higgs unification}

\bibitem{HMOY}
N.\ Haba, S.\ Matsumoto, N.\ Okada and T.\ Yamashita, 
\jnl{\JHEP}{0602}{073}{2006}; 
\reftitle{Effective theoretical approach of Gauge-Higgs unification model and its phenomenological applications}
\\
I.\ Gogoladze, N.\ Okada and Q.\ Shafi, arXiv:0705.3035 [hep-ph]. 
\reftitle{Higgs boson mass from gauge-Higgs unification} 




\bibitem{Pomarol2}
R.\ Contino, Y.\ Nomura and A.\ Pomarol, \jnl{\NPB}{671}{148}{2003}.
\reftitle{Higgs as a holographic pseudo-Goldstone boson}

\bibitem{Agashe2}
K.\ Agashe, R.\ Contino and A.\ Pomarol, 
\jnl{\NPB}{719}{165}{2005}.
\reftitle{The minimal composite Higgs model}


\bibitem{Oda1}
K.\ Oda and A.\ Weiler, \jnl{\PLB}{606}{408}{2005}.
\reftitle{Wilson Lines in Warped Space: Dynamical Symmetry Breaking and Restoration}


\bibitem{HM}
Y.\ Hosotani and M.\ Mabe, \jnl{\PLB}{615}{257}{2005}.
\reftitle{Higgs boson mass and electroweak-gravity hierarchy
from dynamical gauge-Higgs unification in the warped spacetime}


\bibitem{HNSS}
Y.\ Hosotani, S.\ Noda, Y.\ Sakamura and S.\ Shimasaki, 
\jnl{\PRD}{73}{096006}{2006}.
\reftitle{Gauge-Higgs Unification and Quark-Lepton Phenomenology 
in the Warped Spacetime}

\bibitem{YH5}
Y.\ Hosotani,   
in the Proceedings of  CICHEP2, (Cairo, 2006), p.\ 20. (hep-ph/0609035).
\reftitle{Dynamical gauge-Higgs unification}



\bibitem{Carena}
M.\ Carena, E.\ Ponton, J.\ Santiago and C.E.M.\ Wagner,
\jnl{\NPB}{759}{202}{2006};    
\reftitle{Light Kaluza Klein States in Randall-Sundrum Models with Custodial SU(2)}
\jnl{\PRD}{76}{035006}{2007}.  
\reftitle{Electroweak constrains on warped models with custodial symmetry}


\bibitem{SH1} Y.\ Sakamura and Y.\ Hosotani, \jnl{\PLB}{645}{442}{2007}, 
(hep-ph/0607236). 
\reftitle{$WWZ$, $WWH$, and $ZZH$ Couplings in the Dynamical Gauge-Higgs Unification in the Warped Spacetime}

\bibitem{HS2} 
Y.\ Hosotani and  Y.\ Sakamura,  
hep-ph/0703212. 
\reftitle{Anomalous Higgs Couplings in the $SO(5)\times U(1)_{B-L}$ Gauge-Higgs Unification in Warped Spacetime}


\bibitem{Falkowski}
A.\ Falkowski, \jnl{\PRD}{75}{025017}{2007}.
\reftitle{Holographic pseudo-Goldstone boson}

\bibitem{Contino1}
R.\ Contino, T.\ Kramer, M.\ Son and R.\ Sundrum,
\jnl{\JHEP}{0705}{074}{2007}, (hep-ph/0612180).
\reftitle{Warped/composite phenomenology simplified}

\bibitem{GGPR} G.F.\ Giudice, C.\ Grojean, A.\ Pomarol and R.\ Rattazzi, 
\jnl{\JHEP}{0706}{045}{2007},  (hep-ph/0703164). 
\reftitle{The Strongly-Interacting Light Higgs}


\bibitem{PW1}
G.\ Panico and A.\ Wulzer,
\jnl{\JHEP}{0705}{060}{2007}, (hep-th/0703287).
\reftitle{Effective action and holography in 5D gauge theories}

\bibitem{Sakamura}
Y.\ Sakamura,
\jnl{\PRD}{76}{065002}{2007}, (arXiv:0705.1334 [hep-ph]).
\reftitle{Effective theories of gauge-Higgs unification models in warped spacetime}

\bibitem{Falkowski2}
A.\ Falkowski, S.\ Pokorski and J.P.\ Roberts,
arXiv:0705.4653 [hep-ph].
\reftitle{Modelling atrong interactions and longitudinally polarized vector
boson scattering}

\bibitem{Wagner1}
A.D.\ Medina, N.R.\ Shah and C.E.M.\ Wagner,
arXiv:0706.1281 [hep-ph].
\reftitle{Gauge-Higgs unification and radiative electroweak symmetry breaking
in warped extra dimensions}

\bibitem{LMH}
  C.S.\ Lim, N.\ Maru and K.\ Hasegawa, arXiv:hep-th/0605180.
  \reftitle{Six dimensional gauge-Higgs unification with an extra space 
  S$^2$ and the hierarchy problem}

\bibitem{STake}
M.\ Sakamoto and K.\ Takenaga, arXiv:0706.0071 [hep-th], PRD in
press. \reftitle{On Gauge Symmetry Breaking via 
Euclidean Time Component of Gauge Fields}

\bibitem{LM2}
C.S.\ Lim and N.\ Maru,  \jnl{\PLB}{653}{320}{2007}, (arXiv:0706.1397 [hep-ph]).
  \reftitle{Towards A Realistic Grand Gauge-Higgs Unification}

\bibitem{Adachi}
Y.\  Adachi, C.S.\ Lim, N.\ Maru, arXiv:0707.1735 [hep-ph] PRD in press.
\reftitle{Finite anomalous magnetic moment in the gauge-Higgs unification}

\bibitem{RR1}
S.\ Randjbar-Daemi and V.\ Rubakov,  arXiv:0709.1202 [hep-ph].
\reftitle{Towards $Z_2$-protected gauge--Higgs unification}

\bibitem{Oikonomou}
V.K.\ Oikonomou, arXiv:0709.1351 [hep-ph].
\reftitle{Temperature Inversion Symmetry in Gauge-Higgs Unification Models}



\bibitem{Gersdorff}
G.v.\ Gersdorff, N.\ Irges and M.\ Quiros,
\jnl{\NPB}{635}{127}{2002};
\reftitle{Bulk and Brane Rediative effects in Gauge Theories on Orbifolds}
hep-ph/0206029.
\reftitle{Finite Mass Corrections in Orbifol Gauge Theories}





\bibitem{YHscgt2}
Y.\ Hosotani, in the Proceedings of 
{\it ``Dynamical Symmetry Breaking"},  ed. M. Harada and K. Yamawaki 
(Nagoya University, 2004), p.\ 17. (hep-ph/0504272).
\reftitle{Dynamical Gauge Symmetry Breaking by Wilson Lines
in the Electroweak Theory}




\bibitem{MY1}N.\ Maru and T.\ Yamashita,  \jnl{\NPB}{754}{127}{2006},
(hep-ph/0603237).
\reftitle{Two-loop Calculation of Higgs Mass in Gauge-Higgs Unification: 
5D Massless QED Compactified on $S^1$}


\bibitem{YHfinite}
Y.\ Hosotani,  hep-ph/0607064.
\reftitle{All-order Finiteness of the Higgs Boson Mass in the Dynamical
Gauge-Higgs Unification}




\bibitem{Irges}
N.\ Irges and F.\ Knechtli, 
\jnl{\NPB}{719}{121}{2005};  
\reftitle{Non-perturbative definition of five-dimensional gauge theories 
on the $R^4 \times S^1/Z_2$ orbifold}
hep-lat/0604006;  
\reftitle{Non-perturbative mass spectrum of an extra-dimensional orbifold}
\jnl{\NPB}{775}{283}{2007}, (hep-lat/0609045).
\reftitle{Lattice gauge theory approach to spontaneous symmetry breaking from an extra dimension}



\bibitem{LM1}
C.S.\ Lim and N.\ Maru, \jnl{\PRD}{75}{115011}{2007},  (hep-ph/0703017).
\reftitle{Calculable One-Loop Contributions to S and T Parameters 
in the Gauge-Higgs Unification}


\bibitem{Ghilencea1}
D.M.\ Ghilencea, H.M.\ Lee and K.\ Schmidt-Hoberg,
 \jnl{\JHEP}{0608}{009}{2006}.
 \reftitle{Higher derivatives and brane-localised kinetic terms 
 in gauge theories on orbifolds}





\bibitem{Schwinger} J.\ Schwinger, \jnl{\PR}{125}{397}{1962};
   \jnl{\PR}{128}{2425}{1962}.
   
\bibitem{HH} J.E.\ Hetrick and Y.\ Hosotani, \jnl{\PRD}{38}{2621}{1988}.
\reftitle{QED on a Circle}

\end{thebibliography}
\end{document}